\documentclass[12pt,cite,epsf,epsfig]{article}
\usepackage{epsfig}
\title{W-exchange/Annihilation amplitudes in LCSR - $B_d^0 \to D_s^-K^+$ 
as an example }
\author{Namit Mahajan\thanks{E--mail : nmahajan@mri.ernet.in}\\
	{\em Harish-Chandra Research Institute,} \\
	 {\em Chhatnag Road, Jhunsi, Allahabad - 211019, 
India.}\\ 
and\\
{\em High Energy Section,}\\
{\em The Abdus Salam International Centre for Theoretical 
Physics,}\\
{\em Strada Costiera, 11-34014 Trieste, Italy.}}

\def\be{\begin{equation}}
\def\ee{\end{equation}}
\def\bea{\begin{eqnarray}}
\def\eea{\end{eqnarray}}

\begin{document}
%\doublespacing
\maketitle
%\large

\begin{abstract}
We evaluate the W-exchange diagram within the framework of light cone sum 
rules (LCSR), taking $B_d^0 \to D_s^-K^+$ as an example. This decay mode, 
though proceeding via the W-exchange diagram only and therefore expected to be 
highly suppressed, has the branching ratio $3.2 \times 10^{-5}$. We estimate 
the W-exchange amplitude within LCSR, including soft gluon corrections, 
to twist-3 accuracy. The calculation
naturally brings out the features which suggest that such an
amplitude is expected to be 
small in cases when both the final state mesons are light, 
without relying on any kind of an assumption. 
With minor changes, it is also 
possible to have a rough estimate of the exchange/annihilation type
 contributions to other processes like $B \to \pi \pi,~\pi K, KK$. We find
 that though it appears as if the sum rule method yields a fair agreement 
with the observed value, a careful analysis of individual terms
shows that the method, in its
present form, is inadequate to capture the correct physical answer for the case
of heavy final states. 
%{\bf Keywords}: Light cone sum rules, annihilation/W-exchange diagrams
%{\bf PACS}: 
\end{abstract} 

\section{Introduction}
\par There is a compelling need to have an unambiguous quantitative estimate
of the two body hadronic B decay amplitudes. It is now well established that
the naive factorization approach has to be abandoned and
one has to rely on one of the QCD based approaches, like 
QCD factorization (QCDF) \cite{bbns}, perturbative QCD (PQCD) \cite{pqcd}, 
soft collinear 
effective theory (SCET) \cite{scet} or light cone sum rules (LCSR) \cite{lcsr},
to calculate hadronic decay amplitudes. These differ in their treatment
of one or more hadronic quantities/parameters and therefore the results are
quite approach dependent and at times very different.

\par The rare decay $B_d^0 \to D_s^-K^+$ proceeds solely via the W-exchange 
diagram. In general, certain diagrams, which go under the name of 
{\it annihilation (A),  
W-exchange (E)and  penguin-annihilation (PA)} - hereafter
generically referred to as annihilation diagrams or contributions, 
are neglected on the ground
of being much lower in the hierarchy of contributing diagrams. These are the
diagrams where the lighter quark present in the B-meson
 also participates in the weak process as 
opposed to {\it tree (T) or color-suppressed tree (C) 
or penguin (P)} type diagrams. The
annihilation contributions are most often either set to zero or 
become parameters of the model. Since the mode $B_d^0 \to D_s^-K^+$ has only 
the {\it E-}type contribution, it offers a unique opportunity to confront
theoretical predictions against the experimental observations. Also, it
can be used to extract and understand the general structure of such an 
amplitude. This decay mode has been investigated in the past within various
calculational approaches. If factorization holds, then the exchange amplitude
will have the form
\be
E = \frac{G_F}{\sqrt{2}}V_{ud}V^*_{cb}\left(\frac{C_1}{3}+C_2\right)f_B
F_0^{0\to D_sK}
\ee
ie the initial B-meson annihilates into vacuum and the [$D_sK$] pair is
created from the vacuum (represented by the time-like form factor).
Within naive factorization, the time-like form factor is expected to
be small at $q^2=m_B^2$ and also the combination of the Wilson coefficients
appearing is a small number. Therefore, this amplitude is negligible and
thus justifies the neglect in such a scheme.
Using the BSW model \cite{bsw} the branching ratio is predicted to be 
$6.5\times 10^{-8}$
\cite{xing}. In order to have a sizeable branching ratio within such a 
picture, the possible explanation is to consider final state 
rescattering effects which can lift the suppression \cite{fsi}.
An early attempt to predict the branching fraction based upon
PQCD \cite{earlypqcd} yielded a value $(4.7 - 6.6)\times 10^{-6}$ - an 
increase by two orders of magnitude compared to the BSW based prediction. 

\par The decay has been experimentally observed
both by BaBar \cite{babar} and Belle \cite{belle} 
and the observed branching ratio is
\be
BR(B^0_d\to D_s^-K^+) = \Bigg\{\begin{array}{cc}
(3.2\pm 1.0\pm 1.0)\times 10^{-5}& (BaBar)\\
(4.6\pm 1.2\pm 1.3)\times 10^{-5}& (Belle)
\end{array} \label{expt}
\ee
The observed branching ratio is larger than the 
predicted value. It has been re-examined within PQCD \cite{dskpqcd} and 
it is found that 
theoretical prediction matches well with the experimental value. However, any
such calculation requires precise knowledge of heavy meson wave functions, 
which is lacking at the moment. Also, the employed wave functions do not 
satisfy equation of motion constraints. Therefore, once that is taken into
account, some differences may arise.
In \cite{mantry}, within SCET, it is noted that the {\it C-} and 
{\it E-}type 
contributions are of similar size and are both suppressed compared to {\it T-}
contributions and it is expected that the $B_d^0 \to D_s^-K^+$ amplitude
will have a suppression factor of about $3$ compared to  $B_d^0 \to D^0\pi^0$
to account for the experimental numbers. However, a complete calculation
within SCET is still missing. Further, since such 
contributions are free parameters in QCDF or schemes relying on $SU(3)$ 
classification, it is desirable to have an independent check on such results
employing some other method. This is further required by the need to verify
the presence (or absence) of possible large final state interactions in
the channel. It should be noticed that the effect of such rescatterings can
be to enhance the rates and also to bring in extra contributions with 
different CKM elements - thus providing with an opportunity to look for CP 
violation which will unambiguously confirm such rescatterings. However, in the
present case, the dominant rescattering contributions come from $\pi^{+(0)}
D^{-(0)}$ intermediate states. These amplitudes have the same weak phase as 
the E-type contribution and therefore there is no possibility of CP violation.
Therefore, it becomes even more important to have independent checks
to be sure of the results. 
Moreover, an unambiguous estimation of such contributions is necessary
in order to faithfully extract CKM parameters,
since such contributions are present in the decay modes often employed for
extracting CKM angles.

\par Recently, a modified light cone sum rule method has been proposed 
\cite{khodjamirian1} and has been employed to estimate 
emission \cite{khodjamirian1} and penguin contributions \cite{khodjamirian2}, 
both hard and soft as well as factorizable and non-factorizable,
to $B\to\pi\pi$ mode as well as to evaluate 
soft non-factorizable 
contributions in case of $B\to J/\psi K$ \cite{melic},
$B\to D\pi$ \cite{dpi}, $B\to \eta_c K,~\chi_c K$ \cite{charmk},
$B\to$ K +~charmonium \cite{melic2} and $B\to K \pi$ \cite{kpi}. 
It has been shown in these studies 
that the soft gluon contributions are equally important as the hard gluon
ones, if not dominant. However, none of these studies addresses the issue of 
annihilation type diagrams  
within this modified LCSR approach. It is the aim of this study to focus on
the evaluation of such contributions employing LCSR. 

\par We evaluate the W-exchange diagram and work to twist-3 accuracy. 
The calculation explicitly brings out the features which clearly show why
the exchange/annihilation type contributions are generally small.
The leading soft gluon corrections turn out to be
proportional to $q^2=m_K^2$, which vanish in the chiral limit. 
Such soft gluon corrections may not be completely insignificant
numerically due to the presence of large multiplicative factors.
The main aim of the present study is to present an unambiguous and 
clear way of estimating annihilation type contributions,
without having to rely on any dynamical assumptions. 
Using the computed amplitude,
and not relying on general expectations, we argue that the annihilation
type amplitudes are $1/m_B$ suppressed compared to the tree contributions.
However, a careful analysis of individual terms in the final expression
reveals that the sum rule method in its present form is not suitable for final
states containing heavy quark like charm.

\par The paper is organised as follows. In the next section we outline the 
method and introduce the basic correlation function. In Section 3 we present 
the calculation of the correlation function, including the leading soft
gluon corrections. Section 4 contains the numerical results and discussions. 
Conclusions are summarised in Section 5.
%%%%%%%%%%%%%%%%%%%%%%%%%%%%%%%%%%%%%%%%%%%%%%%%%%%%%%%%%%%%%%%%%%%%%%%%%%%
\section{Modified LCSR framework and relevant correlation function} 
\par We are interested in evaluating the amplitude for the decay 
 $B_d^0 \to D_s^-K^+$. It proceeds via the W-exchange diagram, with the 
$s\bar{s}$ pair attached to any of the quark legs via a gluon. 
The effective Hamiltonian relevant for the process is
\be
{\mathcal{H}}_{eff} = \frac{G_F}{\sqrt{2}}V_{ud}V^*_{cb}~
\left[C_1(\mu)O_1(\mu) + C_2(\mu) O_2(\mu)\right] 
\ee
where
\be O_1 = (\bar{b}\Gamma_{\mu}c)(\bar{u}\Gamma^{\mu}d) \hskip 1.5cm
O_2 = (\bar{u}\Gamma_{\mu}c)(\bar{b}\Gamma^{\mu}d)
\ee
where $\Gamma_{\mu} = \gamma_{\mu}(1-\gamma_5)$
The above effective Hamiltonian can be cast into the following
form (suppressing the scale $\mu$ in operators as well as coefficients)
\be
{\mathcal{H}}_{eff} = \frac{G_F}{\sqrt{2}}V_{ud}V^*_{cb}
\left[\left(\frac{C_1}{3}+C_2\right) O_2 + 2 C_1 \tilde{O_2}\right]
\ee
where
\be
\tilde{O_2} = \left(\bar{u}\frac{\lambda^a}{2}\Gamma_{\mu}c\right)
\left(\bar{b}\frac{\lambda^a}{2}\Gamma^{\mu}d\right)
\ee
and $\lambda^a$ are the Gell-Mann matrices with normalization
$Tr(\lambda^a\lambda^b) = 2\delta^{ab}$. The amplitude that we are interested
in is
\be
{\mathcal{A}}(B_d^0\to D_s^-K^+) = \langle D_s^-(p)K^+(q)\vert
{\mathcal{H}}_{eff}\vert B_d^0(p+q)\rangle 
\ee
%%%%%%%%%%%%%%%%%%%%%%%%%
%\vskip 3.0cm
\begin{figure}[ht]
%\vspace*{-1cm}
%\hspace*{-0.5cm}
\centerline{
\epsfxsize=5.0cm\epsfysize=3cm
                      \epsfbox{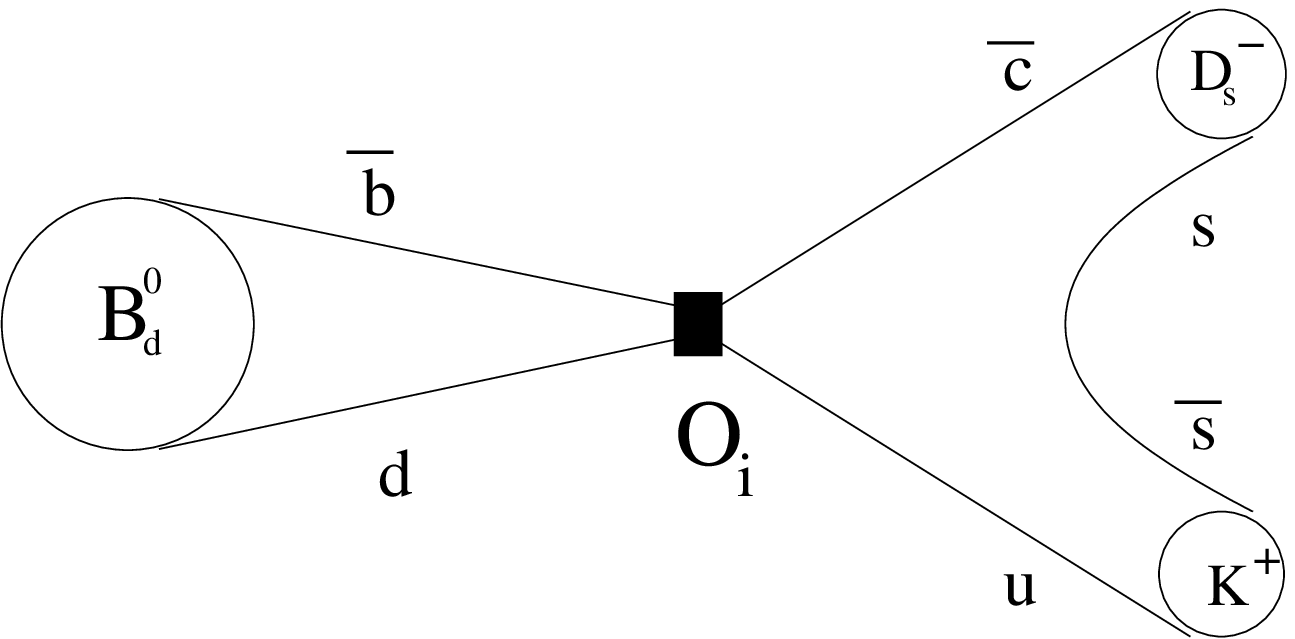}
\hskip 1cm
\epsfxsize=6.0cm\epsfysize=4cm
                      \epsfbox{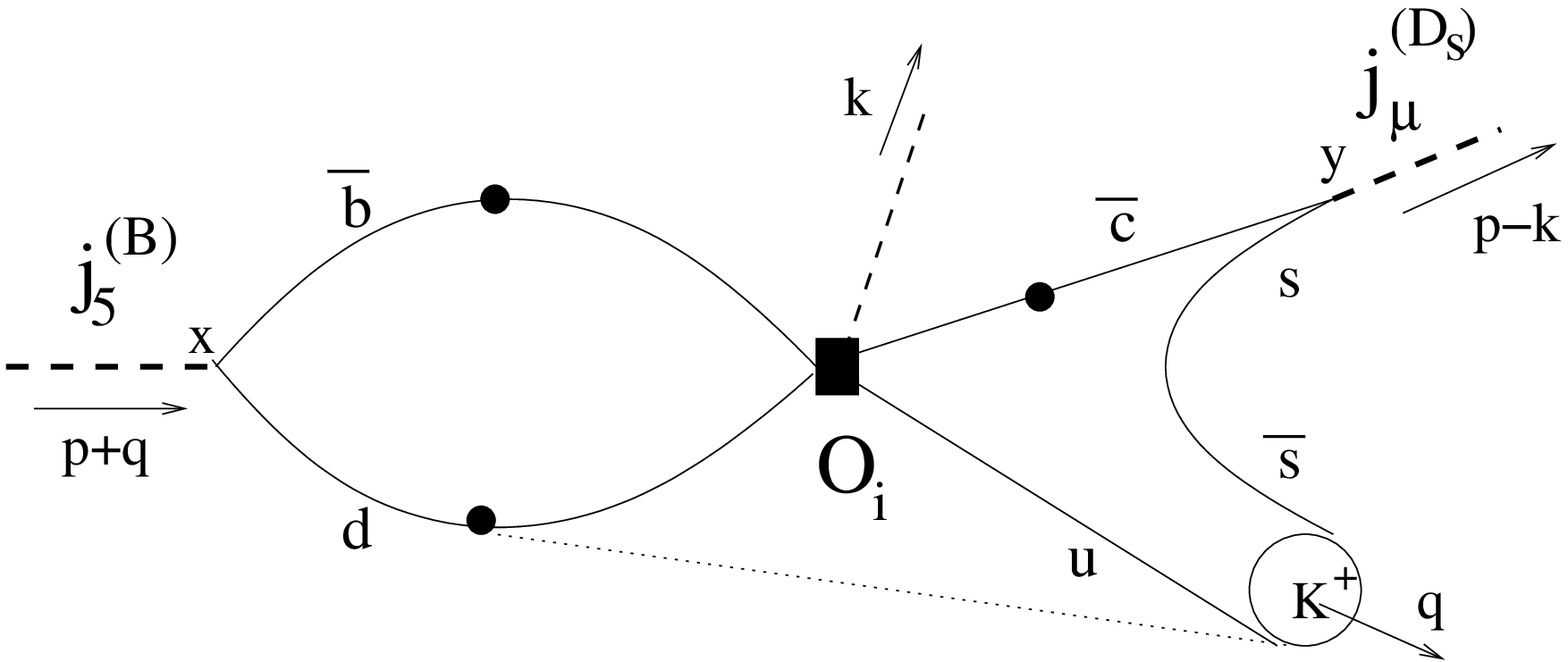}
}
\caption{Schematic representation of the exchange diagram (Left). $O_i$ 
represents one of the operators of the effective Hamiltonian. 
LCSR picture (Right) with B and $D_s$
interpolating currents. $k$ is the artificial four momentum introduced at 
the weak vertex. The thick dot indicates possible lines to which 
a soft gluon (dotted line) from the kaon distribution amplitude can be
 attached.}
\end{figure}
%%%%%%%%%%%%%%%%%%%%%%%%%%
\par The starting point of any LCSR calculation is the specification of the 
relevant vacuum-meson correlation function. In the present case, the 
correlator of interest is the following
\be
F^{(O)}_{\mu}(p,q,k) = -\int d^4x e^{i(p-q)x}\int d^4y e^{i(p-q)y}
\langle K^+(q)\vert T[j_{\mu}^{(D_s)}(y)~O(0)~j_5^{(B)}(x)]\vert
0\rangle \label{cor}
\ee
where $O$ is one of the operators present in the effective Hamiltonian
and $j_{\mu}^{(D_s)}$ and $j_5^{(B)}$ are the interpolating currents for
$D_s$ and B-meson respectively. Explicitly we have,
$O = (\bar{u}^i\Gamma_{\mu}c^j)(\bar{b}^k\Gamma^{\mu}d^l)~A^{ij}A^{kl}$,
where $A^{ij}A^{kl}=\delta^{ij}\delta^{kl}$ for $O=O_2$ and 
$A^{ij}A^{kl}=\left(\frac{\lambda^a}{2}\right)^{ij}
\left(\frac{\lambda^a}{2}\right)^{kl}$ for $O=\tilde{O_2}$ and
$j_5^{(B)}=im_b\bar{d}\gamma_5b$, 
$j_{\mu}^{(D_s)}=\bar{c}\gamma_{\mu}\gamma_5s$.

\par As proposed and explained in \cite{khodjamirian1}, an unphysical,
 artificial four momentum $k$ is introduced in the problem. 
This is done in order to
avoid unwanted contributions to the dispersion relation. This is ensured
with the introduction of the unphysical momentum because now the total momentum
of the final state mesons, $P = (p-k) + q$, 
is different from the momentum of the
initial B-meson, $p + q$. In the final physical matrix elements, no trace of 
this fictitious momentum $k$ remains. The six invariants are chosen to be
$p^2$, $q^2$, $k^2$, $(p-k)^2$, $(p+q)^2$ and $P^2$. The kaon is on-shell.
In the chiral limit,
the kaon can be treated as massless and thus $q^2=0$. We do not assume such
a limit for the time being and shall only consider it towards the end.
The correlation function is evaluated in the deep Euclidean region and then 
analytically continued to the time-like region. The complete kinematical region
in which the  light cone expansion is applicable, relevant to this case is
\[
k^2 = 0 \hskip 1cm q^2 = m_K^2 \hskip 1cm p^2 = m_{D_s}^2
\]
\[
\vert (p+q)\vert^2, \vert (p-k)\vert^2, \vert P\vert^2 >> \Lambda_{QCD}^2
\]
The correlation function, Eq(\ref{cor}), can be expressed in terms of the four
independent tensor structures, namely
\be
F^{(O)}_{\mu}(p,q,k) = F_0^{(O)}~(p-k)_{\mu} + F_1^{(O)}~q_{\mu} + 
F_2^{(O)}~k_{\mu} + 
F_3^{(O)}~\epsilon_{\mu\nu\alpha\beta}p^{\nu}q^{\alpha}k^{\beta}
\ee
For the present case, $F_0^{(O)}$ is the only object of interest to us. 

\par The above correlation function is evaluated to the desired accuracy, both
in the strong coupling, $g_s$, and the kaon 
distribution amplitudes governed by the twist
expansion. The evaluated result is then expressed as a double dispersion
integral with respect to the variables $(p-k)^2$ and $(p+q)^2$ and is
subsequently matched to the corresponding hadronic double dispersion integral.
In the intermediate steps, Borel transformations are applied in both the 
variables and quark-hadron duality is invoked to approximate the excited state
contributions. It is found that the ground state B-meson contribution is
independent of the auxiliary momentum $k$. Thus the final physical matrix
elements do not have any dependence on the fictitious momentum introduced.
The end result of such a matching has the following structure
\bea
{\mathcal{A}}(B_d^0\to D_s^-K^+) &\equiv& [..]~\langle D_s^-(p)K^+(q)\vert
O\vert B_d^0(p+q)\rangle \nonumber \\
&=& [..]~\left(-\frac{i}{\pi^2f_{D_s}f_Bm_B^2}\right)
\int_{m_c^2}^{s_{th}^{D_s}}ds_1 ~e^{(m_{D_s}^2-s_1)/M_1^2}\int_{m_b^2}^{R_b}
ds_2 ~e^{(m_{B}^2-s_2)/M_2^2}~ \nonumber \\
&&\hskip 4cm Im_{s_1}Im_{s_2}F_{0,QCD}^{(O)}(s_1,s_2) \label{hadamp}
\eea
where $[..]$ represents the overall multiplicative factors like CKM elements
and Wilson coefficients relevant to the operator inserted.
%%%%%%%%%%%%%%%%%%%%%%%%%%%%%%%%%%%%%%%%%%%%%%%%%%%%%%%%%%%%%%%%%%%%%%%%%%%%
\section{QCD calculation of correlation function and hadronic matrix elements}
\par The possible contributions to the correlation function up to 
${\mathcal{O}}(g_s)$ can be diagrammatically represented as shown in Fig.1.
These come in two forms - contributions without any soft gluon and with 
one soft gluon connecting the kaon to one of the remaining three quark lines.
The soft gluon contributions are obtained by expanding the relevant 
propagator(s) around the light cone and picking the first non-trivial
terms containing one gluon field \cite{balitsky}:
\bea
S^{jk}(x,y:m) &\equiv& -i\langle 0\vert T[q^j(x)\bar{q}^k(y)\vert 0\rangle \\
&=& \int \frac{d^4k}{2\pi^4}~e^{-i(x-y)k}\Bigg[\frac{\not{k}+m}{k^2-m^2}
\delta^{jk} - \int_0^1 dv g_s G_a^{\mu\nu}(vx + (1-v)y)\left(\frac{\lambda^a}
{2}\right)^{jk} \nonumber \\
&\times& \Bigg\{\frac{1}{2}\frac{\not{k}+m}{(k^2-m^2)^2}\sigma_{\mu\nu}
- \frac{1}{k^2-m^2}v(x-y)_{\mu}\gamma_{\nu}\Bigg\} + 
{\mathcal{O}}(g_s^2)\Bigg] \nonumber
\eea
We discuss and evaluate these contributions below. To this end, we note that
various contributions are labeled according to the operator from the
effective Hamiltonian and the order in strong coupling. We restrict ourselves
to twist-3 order in the kaon distribution amplitudes. Incorporation of higher
twist effects is fairly straightforward, though cumbersome. The contribution
due to insertion of operator $O_2$ is labeled as $F_0^{(O_2)}$. Further, it is
explicitly split into terms according to the order in strong coupling and 
subsequently the contributions are labeled as $F_0^{(O_2,g_s^0)}$ and
$F_0^{(O_2,g_s^1)}$. The former has no gluon line while in the latter, the 
gluon
from the kaon amplitude is hooked on to the charm quark line. 
The last contribution is a sum of two diagrams where the soft 
gluon gets attached to either the bottom or down quark line. These
are the soft gluon contributions arising
due to $\tilde{O_2}$ and are labeled as $F_0^{(\tilde{O_2},g_s^1)}$.
 The calculation is done in the NDR scheme. We note that though all
the diagrams are divergent, the divergent terms vanish on Borel transforming.
Further, we do not show finite terms which disappear after Borel transformation
in either or both the variables or the terms which are not proportional
to the four vector $(p-k)$.  

\par We start by evaluating, $F_0^{(O_2,g_s^0)}$, the leading factorizable
contribution to the correlation function. In the correlation function
Eq(\ref{cor}), use $O=O_2$ and make the relevant Wick contractions. For the
quark-anti-quark T-ordered products, pick the trivial terms. It is
straightforward to evaluate the one-loop Feynman integrals. The end result is
\bea
F_0^{(O_2,g_s^0)} &=& -\frac{f_Km_b^2}{4\pi^2}\int [D\alpha_i]\int_0^1 dx
\frac{1}{[(p'-k)^2-m_c^2]}\Bigg\{[(xQ^2 - q.Q)\phi_K(\alpha_i) \nonumber \\
&-& (x-1)q.(p'-k)\phi_K(\alpha_i) + m_c\mu_K(x-1)\phi_p(\alpha_i)]
\ln(m_b^2 - (1-x)Q^2) \nonumber \\
&+& \frac{m_c\mu_K}{3}(1-x)^2 \frac{q.Q}{m_b^2-(1-x)Q^2}
\phi_{\sigma}(\alpha_i)\Bigg\} \label{hardfact}
\eea
where $[D\alpha_i] = (\Pi_id\alpha_i)\delta (1-\sum_i\alpha_i)$, 
$\mu_K=m_K^2/(m_u+m_s)$, $Q=(p+q)$ and 
$p'=p+\alpha_sq$. The light quark masses are all set to zero.
We choose to label the $\alpha_i$'s by the parton they
correspond to. The different $\phi$'s represent different kaon distribution
amplitudes defined through the following relations \cite{braun}:
\[
\langle K^+(q)\vert\bar{u}(0)\gamma_{\mu}\gamma_5s(y)\vert 0\rangle = 
-iq_{\mu}f_K\int [D\alpha_i]e^{i\alpha_sy.q}\phi_K(\alpha_i)
\]
\be
\langle K^+(q)\vert\bar{u}(0)\gamma_5s(y)\vert 0\rangle = 
-if_K\mu_K\int [D\alpha_i]e^{i\alpha_sy.q}\phi_p(\alpha_i)
\ee
\[
\langle K^+(q)\vert\bar{u}(0)\sigma_{\mu\nu}\gamma_5s(y)\vert 0\rangle = 
i(q_{\mu}y_{\nu}-q_{\nu}y_{\mu})\frac{f_K\mu_K}{6}
\int [D\alpha_i]e^{i\alpha_sy.q}\phi_{\sigma}(\alpha_i)
\]
$\phi_K$ is twist-2 kaon distribution amplitude while $\phi_p$ and 
$\phi_{\sigma}$ are twist-3 distribution amplitudes.

\par Next we consider the ${\mathcal{O}}(g_s)$ soft gluon contribution,
$F_0^{(O_2,g_s^1)}$. This contribution arises when
the soft gluon connects the final state kaon to the charm quark line.  
Following exactly the same procedure  as above, and considering the 
${\mathcal{O}}(g_s)$ term of the propagator, we obtain
\be
F_0^{(O_2,g_s^1)} = 16im_cm_b^2f_{3K}q^2\int_0^1 dv\int [D\alpha_i]
\frac{\phi_{3K}(\alpha_i)}{[(k-p'')^2-m_c^2]^2}\int \frac{d^4k_d}{(2\pi)^4}
\frac{{k_d}_{\mu}}{k_d^2[(k_d-Q)^2-m_b^2]} \label{softfact}
\ee
where $p''=p+(\alpha_s+v\alpha_g)q$ and 
$\phi_{3K}$ is the twist-3 three particle distribution amplitude defined
via the relation
\bea
\langle K^+(q)\vert\bar{u}(0)\sigma^{\alpha\gamma}\gamma_5
{\mathcal{G}}^{\lambda\sigma}(vx)s(y)\vert 0\rangle &=& 
-if_{3K}[(q^{\lambda}q^{\alpha}\eta^{\sigma\gamma} - 
q^{\sigma}q^{\alpha}\eta^{\lambda\gamma}) - 
(q^{\lambda}q^{\gamma}\eta^{\sigma\alpha} - 
q^{\sigma}q^{\gamma}\eta^{\lambda\alpha})] \nonumber \\
&&\int [D\alpha_i]e^{i(\alpha_sy+v\alpha_gx).q}\phi_p(\alpha_i)
\eea
where ${\mathcal{G}}^{\lambda\sigma}(vx) = g_sG_a^{\lambda\sigma}(vx)
\left(\frac{\lambda^a}{2}\right)$. 

\par The last contribution to be evaluated is the soft gluon 
contribution to the correlation function due to $\tilde{O_2}$. 
There are two diagrams to this
piece - with the gluon attaching to the bottom or down quark line. The sum
of these two diagrams yields
\bea
F_0^{(\tilde{O_2},g_s^1)} &=& -2im_cm_b^2f_{3K}q^2\int_0^1 dv\int [D\alpha_i]
\frac{\phi_{3K}(\alpha_i)}{[(k-p')^2-m_c^2]}\nonumber \\
&& \int \frac{d^4k_d}{(2\pi)^4}\frac{{k_d}_{\mu}}{k_d^2[(k_d-Q')^2-m_b^2]}
\Bigg[\frac{1}{k_d^2} + \frac{1}{(k_d-Q')^2-m_b^2}\Bigg] \label{softnonfact}
\eea
where $Q'=p+(1-v\alpha_g)q$.

\par Before proceeding further with the calculation of hadronic matrix 
elements, we would like to discuss some of the features that are evident
from the above computation. In particular, the following is noteworthy.
The soft contributions, Eq(\ref{softfact}) and Eq(\ref{softnonfact}), are
proportional to the momentum squared of the on-shell kaon. If we work in the
chiral limit, where $q^2=m_K^2=0$, then these contributions vanish 
identically. Moreover, both these contributions are proportional to the
charm quark mass. These are two of the most important features of the present
calculation. Imagine computing the annihilation type contribution for the case
of two light mesons in the final state. Then the charm quark mass would have
been replaced by the corresponding light quark mass which would have been set
to zero without introducing any new assumption. This would have implied that 
the soft gluon corrections vanish, even if the corresponding $q^2$ is not
set to zero. Also, some terms drop out from the factorizable contribution, 
Eq(\ref{hardfact}), on similar arguments. We therefore have a natural
explanation for expecting the annihilation type diagrams in most of the cases
to be rather small.

\par  It is worthwhile to mention that setting the charm mass to
zero would imply vanishing of the amplitude arising due to $O_2$ because of 
current conservation. However, the sum rule calculation yields a non vanishing
result and this point must be understood. Consider computing annihilation
amplitude for B-meson decay into light hadrons. In that case, the light quark
masses would have all been set to zero and the only non vanishing contribution
stems from quark mass independent terms in Eq(\ref{hardfact}). However, this
particular contribution will be ${\mathcal{O}}(s_{th}/m_B^2)$, where $s_{th}$
is the corresponding threshold in the light hadron channel. Such a
contribution is to be neglected in the approximation we are working with, and
therefore, the sum rule calculation conforms with the expectation of vanishing
contribution. Let us now consider what happens in our case. We set the charm
mass to zero.  The contribution in the limit of zero charm
mass is ${\mathcal{O}}(s_{th}^{D_s}/m_B^2)\sim (20-30)\%$ for $s_{th}^{D_s}=6$
GeV$^2$. Such a contribution, ${\mathcal{O}}(s_{th}/m_B^2)$, is actually an
artefact of the sum rule method and must be looked upon as an error in the
prediction. In contrast to the case of light final states, this induced error
is quite large. We take this as a hint that the sum rule method, 
in the present form and
within the approximations employed, is too crude to capture the physically
correct answer for the decay into heavy final state(s). However in order to
estimate the numerical error due to such a contribution, we must evaluate the
matrix elements and study individual pieces.

\par Having discussed the important features emerging from the structure of
various contributions to the correlation function, we proceed to the
evaluation of hadronic matrix elements. We choose to
work in the chiral limit ie. we set $q^0=m_k^2=0$. 
In physical terms, this means
that we have neglected ${\mathcal{O}}(q^2/m_B^2)$ terms in the analysis.
Note however that
$\mu_K\neq 0$ in this limit. This implies that the soft gluon corrections
are neglected altogether and we are left with $ F_0^{(O_2,g_s^0)}$ only.
\footnote{We would like to mention a word of caution at this point regarding 
the neglect of the soft gluon contribution due to $\tilde{O_2}$,
$F_0^{(\tilde{O_2},g_s^1)}$. Although, the contribution
is proportional to $q^2$, it comes with factors of $m_c$ and $2C_1$ 
(recall that $C_1\sim 1$) and thus it is possible that the $q^2/m_B^2$
suppression is partially lifted.}
Define two new variables
\be
s_1 = \frac{m_c^2-\alpha_sP^2}{1-\alpha_s} \hskip 1.5cm
s_2 = \frac{m_b^2}{1-x}
\ee 
It is fairly straightforward to make the change of variables in the 
expression for $F_0^{(O_2,g_s^0)}$ to arrive at the following
\bea
F_0^{(O_2,g_s^0)} &=& \frac{f_Km_b^2}{4\pi^2}\int_{m_c^2}^{\infty} ds_1
\int_{m_b^2}^{\infty} ds_2\Bigg[\frac{m_b^2}{s_2^2(s_1-P^2)}\Bigg]
\Bigg[\frac{1}{s_1-(p-k)^2}\Bigg] \\
&\times& \Bigg[\Bigg\{\left[\left(\frac{1}{2}-\frac{m_b^2}{s_2}\right)P^2 - 
\frac{m_b^2s_1}{2s_2}\right]\phi_K(\alpha_s) - \frac{m_c\mu_Km_b^2}{s_2}
\phi_p(\alpha_s)\Bigg\}\ln(s_2-Q^2)\nonumber \\
&+& \left(\frac{1}{2}-\frac{m_b^2}{s_2}\right)\phi_K(\alpha_s) [Q^2\ln(s_2-Q^2)]
+ \frac{m_c\mu_Km_b^2}{6s_2}(s_2-p^2)\phi_{\sigma}(\alpha_s)
\Bigg(\frac{1}{s_2-Q^2}\Bigg)\Bigg] \nonumber
\eea
Borel transforming twice with respect to $(p-k)^2$ and $Q^2=(p+q)^2$
leads to the hadronic matrix element
\bea
{\mathcal{A}}(B_d^0\to D_s^-K^+) &=& \frac{G_F}{\sqrt{2}}
V_{ud}V^*_{cb}\left(\frac{C_1}{3}+C_2\right)
~\langle D_s^-(p)K^+(q)\vert
O_2\vert B_d^0(p+q)\rangle \nonumber \\
&=& \frac{G_F}{\sqrt{2}}
V_{ud}V^*_{cb}\left(\frac{C_1}{3}+C_2\right)
~\left(-\frac{i}{\pi^2f_{D_s}f_Bm_B^2}\right)
\int_{m_c^2}^{s_{th}^{D_s}}ds_1 ~e^{(m_{D_s}^2-s_1)/M_1^2} \nonumber \\
&\times& \int_{m_b^2}^{R_b}ds_2 ~e^{(m_{B}^2-s_2)/M_2^2}~
Im_{s_1}Im_{s_2}F_{0,QCD}^{(O_2,g_s^0)}(s_1,s_2) \label{qcdamp}
\eea
with
\bea
Im_{s_1}Im_{s_2}F_{0,QCD}^{(O_2,g_s^0)}(s_1,s_2)
&=& \left(\frac{f_Km_b^4}{4}\right)
\left[\frac{M_2^2}{s_2^2(s_1-m_B^2)}\right] \\
&\times& \Bigg\{\Bigg[\frac{m_b^2}{s_2}
(m_B^2+M_2^2+\frac{s_1}{2}) - 
\frac{m_B^2+M_2^2}{2}\Bigg]\phi_K(\alpha_s) \nonumber \\
&+& \frac{m_c\mu_Km_b^2}{s_2}\phi_p(\alpha_s) +
\frac{m_c\mu_Km_b^2}{6s_2M_2^2}(s_2-m_{D_s}^2)
~\phi_{\sigma}(\alpha_s)\Bigg\} \nonumber \label{sumrule}
\eea
In the above, we have analytically continued $P^2\to m_B^2$ and 
$p^2\to m_{D_s}^2$. This then completes the LCSR calculation of the 
W-exchange contribution to twist-3 accuracy in the kaon distribution
amplitude in the chiral limit, $q^2=m_K^2=0$. 
Note the absence of any imaginary/absorptive parts indicating that there 
are no rescatterings. Although not shown explicitly, it is easy to conclude
from the structure of the soft corrections that they also do not
introduce any additional phases. 

\par From the structure of the amplitude, Eq(\ref{qcdamp}), it is clear that
we can not write the amplitude in a factorizable form ie as a product of
B-meson decay constant and a form factor. This thus provides
us with a rigorous argument that factorization seizes to hold in 
such amplitudes and thereby offers the justification for naive factorization
failing so badly in predicting the decay rate.

\par The scaling behaviour of various terms of the amplitude in the large $m_B$
limit can be easily obtained. From the above expression of the hadronic
amplitude, it is very much evident that the twist-3 terms are suppressed
by one or two powers of $m_B$ compared to the twist-2 leading term.
We treat the charm quark also as a light quark.
Further, a quick comparison with the heavy mass behaviour studied in
\cite{khodjamirian1} confirms the expectation that the 
annihilation or W-exchange type contributions are ${\mathcal{O}}(1/m_B)$
of the tree/emission type contributions. The soft gluon
contributions are $1/m_B^2$ suppressed, which further justifies
their neglect.  
%%%%%%%%%%%%%%%%%%%%%%%%%%%%%%%%%%%%%%%%%%%%%%%%%%%%%%%%%%%%%%%%%%%%%%%%%%
\section{Numerical estimates}
\par We begin by specifying the input parameters. The NLO values for Wilson 
coefficients in NDR \cite{buras} are $C_1=1.082$ and $C_2=-0.1.85$. 
The absolute values of the CKM elements are taken to be 
$\vert V_{cb}\vert=0.043$ and $\vert V_{ud}\vert=0.974$. For the form of
various distribution amplitudes and other sum rule parameters we rely on
\cite{braun,khoru1,khoru2}. We only specify the central values of various 
parameters
that we have taken into account for the numerical estimation: 
$m_{D_s}=1.968$ GeV,
$m_c=1.3$ GeV, $f_{D_s}=0.22$ GeV, $s_{th}^{D_s}=6$ GeV$^2$, 
$M_1^2=1.5$ GeV$^2$,
$m_B=5.28$ GeV, $m_b=4.8$ GeV, $f_B=0.18$ GeV, $\bar{R}=s_{th}^B=35$ GeV$^2$,
$M_2^2=10$ GeV$^2$, $f_K=0.16$ GeV, $f_{3K}=0.0026$ GeV$^2$, 
$\mu_b=\sqrt{m_B^2-m_b^2}=2.4$ GeV, $\mu_K(\mu_b)=2.5$ GeV.
To have an estimate of the variation of the results with the sum rule
parameters, we check for the variation in the following interval:
$0.8 \leq M_1 \leq 1.5$ and $2.8 \leq M_2 \leq 3.5$.

\par For the two body decay into two pseudoscalars, the decay rate is
simply given as
\be
\Gamma(B\to P_1 P_2) = \Bigg(\frac{1}{16\pi m_B}\Bigg)\vert
{\mathcal{A}}(B\to P_1P_2)\vert^2
\lambda^{1/2}\Bigg(1,\frac{m_{M_1}^2}{m_B^2},\frac{m_{M_2}^2}{m_B^2}\Bigg)
\ee
where $\lambda(a,b,c)=a^2+b^2+c^2-2ab-2ac-2bc$. Using this expression, with
$m_K^2=0$, the absolute value of the amplitude required by the experimental
data Eq(\ref{expt}) is (in GeV) (we quote the numbers corresponding to
the central values of BaBar and Belle results)
\be
\vert {\mathcal{A}}(B_d^0 \to D_s^-K^+)\vert_{expt} \sim (6.5 - 9)
\times 10^{-8} \label{expamp}
\ee
Using the input values of the parameters listed above, to the twist-3
accuracy, we obtain the following value for the amplitude
\be
\vert {\mathcal{A}}(B_d^0 \to D_s^-K^+)\vert_{Tw3, LCSR} = 
\left(3.6^{+2.1}_{-1.9}\right) 
\times 10^{-8} \label{lcsramp}
\ee  
where the upper and lower values correspond to the two extreme values for the
sum rule parameters, $M_1$ and $M_2$.
The central value is about half the experimental value. 
However, we must remember that the
present analysis only takes into account terms up to twist-3 in the kaon
distribution amplitude. We can expect that twist-4 contributions will
partially take care of the discrepancy. Also, recall that some contribution 
is expected from the soft gluon corrections, even at twist-3 level,
because the $q^2/m_B^2$ suppression is lifted due to the large multiplicative
factors, $2C_1m_c$. Therefore, it is tempting to conclude that the experimental
values will be completely saturated once these extra contributions are 
accounted for.

\par However, in view of the discussion related to the charm mass independent
terms in the previous section, it is very important to investigate the
individual contributions separately. Let us recall some of the important
points related to the individual contributions. There are two type of terms -
charm mass independent (twist-2), ${\mathcal{A}}^{ind} 
{\mathcal{O}}(s_{th}^{D_s}/m_B^2)\sim (20-30)\%$, 
and terms proportional to the
charm quark mass (twist-3), ${\mathcal{A}}^{dep}$. A quick check on the
relative size of the two contributions reveals that the bulk of contribution
arises from the charm mass independent terms. Although, this agrees with the
expected scaling behaviour of the individual terms, this observation implies
that the sum rule method yields a much larger theoretical error than we had
anticipated. This thus forces us to conclude that the method is not suitable
for the decay of B-mesons into final states containing a heavy quark like
charm. Further, we do not expect the higher twist contributions to change the
picture drastically as the dominant term in the present case is twist-2, which
will continue to dominate even in the presence of twist-4 or higher terms.
In Table 1 we summarize the values of the two contributions for
different choices of the sum rule parameters - $M_1$ and $M_2$.
\begin{table}[ht]
    \caption{Individual contributions to the amplitude for different choices
    of $M_1$ and $M_2$. The middle row corresponds to the central value of the
    parameters.}
    \begin{center}

    \begin{tabular}{|c|c|c|}
	\hline
$(M_1,M_2)$ (GeV)&${\mathcal{A}}^{ind}$  & ${\mathcal{A}}^{dep}$ \\
	\hline
(0.8,2.8)&$3.16 \times 10^{-8}$&$-1.4 \times 10^{-8}$\\
(1.22,3.16)&$4.26 \times 10^{-8}$&$-6.9 \times 10^{-9}$\\
(1.5,3.5)&$6.53 \times 10^{-8}$&$-8.1 \times 10^{-9}$\\
\hline
\end{tabular}
\end{center}
\end{table}
\par From the table it is clear that the two contributions come with opposite
sign and it is only for the lower values of $M_1$ and $M_2$ that the two are
almost equal in magnitude. Also clear from the table is the fact that the
expected error, an artefact of the method, overwhelms the total
contribution. Although we had anticipated that the error in the present case
is going to be large, the values in the table are at complete variance with 
our naive expectations. 
Therefore, in the present form, the sum rule method is not
reliable for the case of final states involving a heavy quark.
%%%%%%%%%%%%%%%%%%%%%%%%%%%%%%%%%%%%%%%%%%%%%%%%%%%%%%%%%%%%%%%%%%%%%%%%%%
\section{Results and discussion}
\par Employing the modified light cone sum rule method 
\cite{khodjamirian1}, we have computed the hadronic matrix element
for the rare decay $B_d^0 \to D_s^-K^+$. This decay channel falls under
the very special class of decays which receive contribution only from
annihilation/W-exchange diagrams. Such diagrams lie way below in the 
usual hierarchy of diagrams/topologies and are generally neglected.
In such a situation, the only way to have a sizeable branching ratio 
is to expect large final state interactions. Moreover, annihilation 
type diagrams contribute to many other channels like $\pi\pi$ or $\pi K$,
to mention the obvious. It is rather difficult to cleanly extract various
standard model parameters if a precise knowledge of these amplitudes is
not known. 

\par Motivated by all these factors, we have evaluated the W-exchange
diagram to twist-3 accuracy within the LCSR method. We find that
naively it appears that the sum rule method is more or less successful in
explaining the observed branching ratio, while a careful analysis of individual
terms leads us to the conclusion that such an expectation is completely
wrong. Within the accuracy of the sum rule method, we neglect
${\mathcal{O}}(s_{th}/m_B^2)$ terms. However, in the present case such terms
are already ${\mathcal{O}}(20\%)$. Therefore a large error is expected.
We find that bulk of the
contribution to the amplitude actually arises from the terms that should be
viewed as theoretical error. We therefore take this as a clear indication of
the fact that the sum rule method, in the present form and within the adopted
accuracy, is unsuitable for explaining the decay amplitudes when there is a
massive quark in the final state, though the method seems to work for light
final states. The only consistent way out would be to try
to extend the method beyond the adopted approximations and carefully
investigate whether such an extension leads to more meaningful results and
predictions. This would require systematically incorporating, at least the
leading, ${\mathcal{O}}(s_{th}/m_B^2)$ terms. However, it is not clear if such
a modification will be easy to implement or if the whole approach is to be
changed. 

\par In view of the above discussion, it
 is tempting to use the above results to have a crude estimate
of annihilation type diagrams for the decay to light mesons, like
$B\to \pi\pi$. This can be easily achieved by appropriately
changing various parameters.  A crude estimate
for the ratio of the annihilation type amplitudes to the
factorizable amplitude in the $\pi\pi$ channel turns out 
to be at the percent level at best. However, this must include the higher
 order effects also and should be investigated in detail.
The annihilation type diagrams may contribute
sub-dominantly to the decay rate, but still can have significant impact
on CP asymmetries. A systematic study is thus called for in such cases and will
be reported elsewhere.

\par We have argued above that the method can not be trusted for explaining
the annihilation diagrams in case of final state(s) involving a heavy
quark. However, also clear from the above discussion is the fact that the
method is well suited for light final states. It may be worthwhile to present
the summary of the main features which emerge from the analytic structure of
the expressions. For this discussion we do not bother about the numerical
values of the individual pieces as we hope that such a discussion is useful in
understanding the basic structure of a generic annihilation diagram - of course
we keep in mind all the discussion and conclusions relevant for a massive final
state. 
\begin{enumerate}
\item[$\bullet$]Factorizable contribution to twist-3 order is proportional to
  the final state quark mass (the quark mass independent term is an error due
  to the method itself). Such a contribution thus vanishes for light final
  states and one can hope that the twist-4 contributions will yield the
  leading non-vanishing contribution.
\item[$\bullet$]The soft gluon non-factorizable contributions
turn out to be proportional to $q^2$, the mass squared
of the meson described by the distribution amplitudes and mass 
of the (anti-)quark
emerging out of the weak vertex from the bottom (anti-)quark. In the present
case, the proportionality factor thus is $m_cq^2$, with $q^2=m_K^2$.
Such contributions are expected to vanish in the chiral limit. However,
for cases where the final state contains a heavy and a light meson,
these contributions can become significant as there is an additional
enhancement factor, $2C_1$, yielding a net enhancement factor of $2m_cC_1$
for some of the soft gluon contributions.
\item[$\bullet$]The annihilation/W-exchange amplitudes are 
${\mathcal{O}}(1/m_B)$ compared to the tree/emission type amplitudes. 
Further, the twist-3 contributions are suppressed by additional
powers of the large mass in comparison to the twist-2 terms.
\item[$\bullet$]It is not possible to write the amplitude in a factorized form.
\item[$\bullet$]The amplitudes are all real to twist-3 accuracy,
implying absence of rescattering. 
\end{enumerate}

\par To conclude, we have described the evaluation of annihilation type
amplitudes within the framework of light cone sum rules and applied
to the case of $B_d^0 \to D_s^-K^+$. Our results indicate that the modified
sum rule method is not in a form which can be applied to the case of B-meson
decaying into a heavy final state. The numerical results clearly show that the
dominant contribution stems from terms that are an artefact of the sum rule
method and strictly speaking, should be considered a part of the theoretical
error. Only for some particular choices of the sum rule parameters, the
different contributions approach each other in magnitude. Although, the
present study shows that the method fails when trying to explain the mode
 $B_d^0 \to D_s^-K^+$ (similar conclusion will hold for any other heavy
state), the calculation brings out some generic features of a typical
annihilation type diagram. The present computation, in principle, 
completes the computation
of all types of quark level diagrams within LCSR. 
It is hoped that with straightforward
modifications and improvements, the results of this study and the ones already
existing can be combined to obtain a clear and consistent picture
of the two body hadronic B decays. 

%%%%%%%%%%%%%%%%%%%%%%%%%%%%%%%%%%%%%%%%%%%%%%%%%%%%%%%%%%%%%%%%%%%%
\vskip 1cm
{\bf Acknowledgements} I would like to thank the High Energy Section, 
The Abdus Salam International Centre for Theoretical Physics, Trieste, Italy, 
for hospitality during the final stages of the work.
%%%%%%%%%%%%%%%%%%%%%%%%%%%%
%\pagebreak

%%%%%%%%%%%%%%%%%%%%%%%%%%%%%%%%%%%%%%%%%%%%%%%%%%%%%
\end{document}